\documentclass[a4paper,10pt,onecolumn]{article}
\usepackage{amsmath}  
\usepackage{a4wide}                                                          

\usepackage{epsfig} 
\usepackage{mathptmx} 
\usepackage{times} 
\usepackage{amsmath,amssymb}
\usepackage{color}
\newtheorem{assumption}{Assumption}

\newtheorem{lemma}{Lemma}
\newtheorem{remark}{Remark}
\newtheorem{theorem}{Theorem}
\newtheorem{definition}{Definition}

\title{Distributed $H_{\infty}$ Tracking Control for Discrete-Time
  Multi-Agent Systems with a High-Dimensional Leader\\[1cm]
\normalsize
To appear in the Proceedings of the 52nd IEEE Conference on Decision and
Control, 2013, Florence, Italy}

\author{Guanghui Wen and Valery Ugrinovskii
\thanks{This research was supported under Australian Research Council's Discovery
Projects funding scheme (Project number DP120102152). Part of this work was
carried out during the second author's visit to the Australian National
University.}
\thanks{G. Wen and V. Ugrinovskii are with the School of Engineering and
  Information Technology, UNSW Canberra at the Australian Defence Force Academy,
  Canberra, ACT 2600, Australia
        {\tt\small wenguanghui@gmail.com;v.ugrinovskii@gmail.com}.}%
}
\begin{document}

\maketitle
\thispagestyle{empty}
\pagestyle{empty}

\begin{abstract}
This paper considers the distributed $H_{\infty}$ leader-following tracking
problem for a class of discrete-time multi-agent systems with a
high-dimensional dynamic leader. It is
assumed that output information about the leader is only available to
designated followers, and the dynamics of the followers are subject to perturbations.
To achieve distributed $H_{\infty}$ leader-following tracking, a
new class of control protocols is proposed which is based on the
feedback from the nearest neighbors as well as a
 distributed state estimator. Under the
assumptions that dynamics of the leader are detectable
and the communication topology contains a directed spanning tree,
sufficient conditions are obtained that enable all followers to track the
leader while achieving a desired $H_{\infty}$ leader-following tracking
performance. Numerical simulations illustrate the
effectiveness of the theoretical analysis.
\end{abstract}

\section{Introduction}
Distributed control of multi-agent
systems has been receiving a great deal of attention in the recent
literature due to its broad applications in a number of areas; e.g.,
see~\cite{Saber-2007}-\cite{Ugrinovskii-Automatica}. The leader-follower tracking problem
represents a
particular class of distributed control problems which is concerned with
the design of control protocols for each agent based only on the
information from the nearest neighbours, with the aim to guarantee that
states of all followers converge to that of a dynamic
leader; e.g, see~\cite{Hong-2006}.

 This paper considers the leader-following tracking problem
 for a class of discrete-time linear multi-agent systems with a high-dimensional
 leader and undirected communications between followers. Closely related
 work includes 
 \cite{Cao-Automatica-2009}--\cite{Zhongkui-DCDB}.
These references exemplify a common
trend in the existing literature, which has a significant focus on
consensus of discrete-time multi-agent systems with first or second-order
integrator dynamics. In contrast, consensus tracking for discrete-time
multi-agent systems with identical linear higher-order node dynamics was studied in
\cite{KeyouRNC,Zhongkui-DCDB}. The work in \cite{KeyouRNC,Zhongkui-DCDB} covers the results on consensus tracking
for multi-agent systems consisting of first- and second-order integrator dynamics as special cases, respectively.
Also, in the majority of the existing papers on this topic, including some
of the previously mentioned references,
the leader and all the followers are assumed to have identical dynamics models.
This assumption allows one to directly analyze dynamics of the tracking error
arising in the corresponding multi-agent networks consisting of closed-loop
agent systems. On the contrary, the case where dynamics of the leader and
those of the followers have different models (e.g., are described by
state-space equations of different order) has not received as much attention.

In this paper, we focus on the case where dynamics of the leader are more
complex than those of the followers. Therefore, the existing theoretical
approaches for analyzing leader-following tracking problems which have been
developed for networks of identical agents cannot be directly applied in
this case. Furthermore, the state of the leader which evolves independently of
the followers is not directly measurable by all of the followers. It is
assumed that only a partial information about the state of the
leader can be sensed by a small group of followers which are subject to
uncertainty. Thus, the multi-agent system under consideration is more general
and contains some other commonly studied classes of leader-following
multi-agent systems such as, e.g., the multi-agent systems in
\cite{Zhongkui-DCDB}, as special cases. The control goal here is to design
a tracking protocol for each agent such that the closed-loop system of
agents achieves a desired level of $H_{\infty}$ leader-following tracking
performance. Note that the proposed problem of distributed $H_{\infty}$
leader-following tracking for multi-agent systems with a high-dimensional
leader is meaningful in a number of practical applications such as the
design of distributed sensor networks \cite{Fang-WCICA2012} where
dynamics of the leader are more complex than those of the followers.

The fact that the followers can only sense partial information about the
state of the leader prompts us to introduce a dynamic protocol where a
local controller together with a neighbor-based state observer is
designed for each follower. In the present framework, the state observer
embedded in the followers performs the task of estimating the unmeasurable
states of the leader in a distributed way. Under the
assumptions that the leader is detectable and the communication
topology contains a directed spanning tree, we propose a procedure
for  the design of a tracking protocol which involves a solution to a
modified algebraic Riccati equation. The analysis of distributed
$H_{\infty}$ leader-following tracking performance of this protocol when applied
to a multi-agent system with a high-dimensional leader is then
presented.

The protocol design to achieve a pre-specified level of $H_{\infty}$
leader-following tracking performance for a system of agents whose
dimension are different form that of the leader is the main
contribution of this paper. We note that dynamic protocols have been
considered in a number of recent papers. For example, a dynamic protocol
for synchronization of multiagent systems has been proposed recently in
\cite{TTM-2013}. Similar to this paper, the analysis in \cite{TTM-2013} is
based on the reduction of the problem to the analysis of $N$ decoupled
systems. However, in contrast to our paper, robust performance issues are
not considered in \cite{TTM-2013}.  It is also worth noting that there is
another possible way to solve tacking problems under partial information
about the leader, by using the distributed output regulation
theory and internal model principle \cite{Hong-IJRNC2012}.

The remainder of this paper is organized as follows. In
Section II, some preliminaries from the graph theory and the problem formulation are
given. In Section III, the main results are presented. A numerical example
and simulations to illustrate our theoretical analysis are provided in
Section IV. Section V concludes the paper.

\par
\paragraph*{Notation} Let $\mathbb{R}^{n\times n}$ and $\mathbb{C}^{n\times
  n}$ be the sets of $n\times n$ real
matrices and complex matrices, respectively. Let $\mathbb{N}$ and
$\ell_2^{n}$ be, respectively, the sets of natural numbers and the
$n$-dimensional real square summable
functions. If not explicitly stated, all matrices are assumed to have
compatible dimensions. The superscripts $T$ and $H$ denote the transpose
and the Hermitian adjoint of a matrix, respectively. A matrix $U\in
\mathbb{C}^{n\times n}$ is a unitary matrix if $U^{H}U =UU^{H} =
I_n$. $\mathrm{diag}(a_{1},a_{2},\cdots,a_{n})$ represents a diagonal
matrix with $a_{i}$, $i=1,2,\cdots,n$ on its diagonal. The notation
$\textbf{1}_{n}\in \mathbb{R}^{n}$ denotes the vector whose elements are
equal to $1$. Let $O_{n}$ and $I_{n}$ be the $n\times n$ zero and identity
matrices, respectively.  A square matrix is said to be Schur stable if the
magnitude of all of
its eigenvalues is less than $1$.
The symbols $\otimes$ and $\|\cdot\|$ denote, respectively, the Kronecker
product and the Euclidian norm.

\section{Preliminaries and the problem formulation}

\subsection{Preliminaries}

Let $\mathcal{G}(\mathcal{V},\mathcal{E},\mathcal{A})$ be a directed
graph with a set of nodes $\mathcal{V} = \{\upsilon_1, \upsilon_2,
\cdots, \upsilon_N\}$, a set of directed edges $\mathcal{E} \subseteq
\mathcal{V}\times \mathcal{V}$, and a weighted adjacency matrix
$\mathcal{A} = [a_{ij}]_{N\times N}$. We will use the
simplified notation
$\mathcal{G}$ for the graph when this causes no confusion.
A directed edge $e_{ij}$ is associated with an ordered
pair of nodes $(\upsilon_j , \upsilon_i)$, where $\upsilon_j$ and
$\upsilon_i$ are called the parent and child nodes, respectively,
and $e_{ij} \in \mathcal{E}$ if and only if $a_{ij}> 0$. Furthermore,
self-loops are not allowed, i.e.,
$a_{ii}=0$ for all $i=1,2,\cdots,N$. A directed
path from node $\upsilon_i$ to $\upsilon_j$ is an ordered sequence of edges,
$\{(\upsilon_i, \upsilon_{k_1}), (\upsilon_{k_1}, \upsilon_{k_2}),
\ldots, (\upsilon_{k_l}, \upsilon_j)\}\subseteq \mathcal{E}$, with distinct nodes
$\upsilon_{k_m}, m = 1, 2, \cdots, l$.  A directed tree is a directed graph
such that (a)
its every node $\upsilon_{k}$ $(k\neq r)$, except for the root node
$\upsilon_{r}$, has exactly one parent, and (b) there exists a unique
directed path from $\upsilon_{r}$ to each node $\upsilon_{k}$ $(k\neq r)$.
A directed spanning tree of
$\mathcal{G}$ is a directed tree that has the same node set  $\mathcal{V}$
and whose edge set is a subset of $\mathcal{E}$. $\mathcal{G}$ will reduce to an undirected graph  if and only $a_{ij}=a_{ji}$, for all $i,j=1,2,\cdots,N$. A matrix $D=[d_{ij}]_{N\times N}$ is called a row-stochastic
matrix associated with the graph $\mathcal{G}$, if the following properties
hold:
$d_{ii} > 0$; $d_{ij}>0$
if $(j,\, i)\in \mathcal{E}$, and $d_{ij}=0$ otherwise; and
$\sum_{j=1}^{N}d_{ij}=1$, for all $i=1,2,\cdots,N$.

\begin{lemma}[\cite{Ren2005}]\label{LemmaRowSto}
For any row-stochastic matrix $D$ associated with the graph $\mathcal{G}$,
1 is an eigenvalue of $D$, and all other eigenvalues of $D$ lie in the
open unit disk. Furthermore, 1 is a simple eigenvalue of $D$ if and
only if the graph $\mathcal{G}$ contains a directed spanning tree.
\end{lemma}

\subsection{The multi-agent system}

Consider a group of $N$ agents indexed by $1,2,\cdots,N$. Without loss of
generality, it is assumed that the agent labeled $1$ is the leader,
whose dynamics are governed by the following equations
\begin{equation}\label{LeaderDynamics}
\begin{aligned}
\Theta(k+1)&=\hat{A}\Theta(k), \\
y(k+1)&=\hat{C}\Theta(k+1),\quad k\in \mathbb{N}.
\end{aligned}
\end{equation}
The vector $\Theta(k)\in \mathbb{R}^{nm_{0}}$ represents the state of the
leader at time instant $k$, $m_{0}$ and
$n$ are two positive integers. This vector is assumed to be partitioned as
$\Theta(k)=(\theta_{1}(k)^{T},\theta_{2}(k)^{T},\cdots,\theta_{n}(k)^{T})^{T}$,
where
$\theta_{i}(k)\in \mathbb{R}^{m_{0}}$, $i=1,2,\cdots,n$. Accordingly, the state matrix
$\hat{A}=[\hat{a}_{ij}]$ is in $\mathbb{R}^{nm_{0} \times nm_{0}}$, and
admits a compatible partition of the following form:
\begin{equation}\label{partition}
\hat{A}=\left(\begin{array}{cccc}
\hat{A}_{11}&\hat{A}_{12}&\cdots&\hat{A}_{1n}\\
\hat{A}_{21}&\hat{A}_{22}&\cdots&\hat{A}_{2n}\\
\vdots&\vdots&\ddots&\vdots\\
\hat{A}_{n1}&\hat{A}_{n2}&\cdots&\hat{A}_{nn}
\end{array}\right),
\end{equation}
where $\hat{A}_{ij}\in \mathbb{R}^{m_{0}\times m_{0}}$, $i,j=1,2,\cdots,n$.
The vector $y(k+1)$ represents the output information about the leader that
is made available to the followers for sensing  at time instant
$k+1$. However, as will be seen later, it will be assumed that only a
partial information about this vector is sensed by some of the
followers. Therefore, without loss of generality it is assumed that
$\hat{C}=(O_{m_{0}},\cdots,O_{m_{0}},\,I_{m_{0}})\in 
\mathbb{R}^{m_{0}\times nm_{0}}$, i.e., $y(k)=\theta_{n}(k)$, for $k\in
\mathbb{N}$.
The above model for the leader reflects the common situation where the
leader of the group plays the role of a command generator providing
reference states to be tracked by the followers. Therefore, it is natural
to assume that the state of the leader evolves in accordance with its
intrinsic nominal model and is not influenced by the followers.

The dynamics of the followers, labeled as $i$, $i=2,3,\cdots,N$, are
described by the 
state equations
\begin{equation}\label{FollowerDynamics}
x_{i}(k+1)=Ax_{i}(k)+u_{i}(k)+B_{\omega}\omega_{i}(k), \quad k\in \mathbb{N},
\end{equation}
where $A\in \mathbb{R}^{m_{0}\times m_{0}}$ and $B_{\omega}\in \mathbb{R}^{m_{0}\times m_{\omega}}$ are constant matrices,
 $u_{i}(k)\in \mathbb{R}^{m_{0}}$ is the control input to be designed, and
 $\omega_{i}(k)\in \ell_2^{m_{\omega}}$ is a disturbance input.
As mentioned, we assume that some of the followers are able to sense the
output of the leader. In a general form, the output $\hat{y}_{i}(k)$ of the
 leader sensed by the follower $i$, $2\leq i \leq N$, is expressed as
\begin{equation}\label{OutputInformation}
\hat{y}_{i}(k)=c_{i}Ey(k),
\end{equation}
where $c_{i}\geq 0$, and $c_{i}>0$ if and only if the leader is a neighbor
of follower $i$. The matrix $E\in \mathbb{R}^{m_{y}\times m_{0}}$ characterizes
the protocol for information exchange in the network, and
$m_{y}\leq m_{0}$. This reflects the situation where only a partial
information  about the leader is made available to selected followers.
Clearly, if the agent $i$ is not connected to the
leader, then $\hat{y}_{i}(k)=0$.

The present framework has several features that distinguish it from similar
problems considered in the literature. Firstly, while we assume
that the state matrix $\hat{A}$ is known to the
followers, though the initial condition $\Theta(0)$ is unknown, the leader
and the followers have significantly different models in that the states of
the leader and the followers have different dimensions, and are also
governed by different state matrices. Secondly, the followers employ different
output matrices $c_iE$; the scaling parameters $c_i$ may reflect
differences in the strength of the leader's signal received by the
followers positioned at a different distance from the leader. Finally,
dynamics of the followers are subject to uncertain perturbations. As is
well known, within the $H_\infty$
framework, such perturbations are often associated with unmodeled uncertain
dynamics, and can be used to account for imperfections in the followers' models.

In regard to communications between the followers, define the information
output of follower $i$ to be
\[
y_{i}(k)=Ex_{i}(k), \quad 2\leq i \leq N,
\]
where  $E\in \mathbb{R}^{m_{y}\times m_{0}}$ is the matrix from equation
(\ref{OutputInformation}). We will assume that each follower $i$ can only
use for control the information outputs of its neighbours relative its own
output, $y_{i}(k)-y_{j}(k)$.
As mentioned previously, the matrix $E$
characterizes the protocol for information exchange in the network. The
matrix $E$ is in general a rectangular matrix; this allows for the vectors
$y_{i}(k)-y_{j}(k)$ exchanged between the neighboring followers $i$ and $j$
to have a lower dimension than their states. Note that all agents are
assumed to use the same matrix $E$. As discussed in \cite{U7b}, in certain
applications, using a common communication matrix by all followers is not a
significant limitation. 

For the notational convenience, let%
\footnote{The assumption is not restrictive and is made for simplicity. We can
  dispense with this assumption by introducing a modified adjacency matrix
  in which $a_{i1}$ is replaced with $a_{i1}c_{i}$, and then using this
  modified matrix instead of $\mathcal{A}$ in the subsequent derivations.} 
$a_{i1}=c_{i}$ and introduce the
relative information available to follower $i$, $2\leq i \leq N$, given by
\begin{eqnarray}\label{RelativeInformation}
\varepsilon_{i}(k)&=&
\frac{1}{\kappa}\!\bigg[\sum_{j=2}^{N}\!a_{ij}(y_{i}(k){-}y_{j}(k))
+\left(c_i y_{i}(k)-\hat y_i(k)\right)\!\bigg],\! \nonumber \\
&=&\frac{1}{\kappa}\!\bigg[\sum_{j=2}^{N}\!a_{ij}E(x_{i}(k){-}x_{j}(k)){+}a_{i1}E\left(x_{i}(k){-}\theta_{n}(k)\right)\!\bigg],\!
\quad
\end{eqnarray}
where $\kappa=\kappa_{0}+h$,
$\kappa_{0}=\mathrm{max}_{i=2,3,\cdots,N}\{\sum_{s=1}^{N}a_{is}\}$, 
and $h$ is a given positive constant. The factor $1/{\kappa}$ on the right
hand side of
(\ref{RelativeInformation}) is a weighting factor which scales the actual
communication weights between follower $i$ and its neighbors into
positive scalars within the interval $(0,1)$, for each
$i=2,3,\cdots,N$. Note that such scaling
technique is commonly used in  consensus problems for discrete-time multi-agent
systems \cite{Zhongkui-DCDB,Ren2005,ZhongkuiIJC}.

Before closing this section, we state the standing assumptions about the
structure of the system communication topology. The first assumption is
concerned with interactions between the followers, while the second
assumption describes the communication topology between the leader and the
rest of the network.   

\begin{assumption}\label{AssumptionUndirected}
The adjacency matrix of graph $\mathcal{G}$, $\mathcal{A}=[a_{ij}]_{N\times
  N}$ has the property that for all $i,j=2,3,\cdots,N$, $a_{ij}=a_{ji}$.
\end{assumption}

\begin{remark}
Note that Assumption \ref{AssumptionUndirected} indicates that the subgraph
describing the communication topology between the followers is
undirected. However, this subgraph is not required to be connected in the
present framework. 
\end{remark}

\begin{assumption}\label{AssumptionSpanningTree}
The communication topology graph $\mathcal{G}$ contains a directed spanning
tree with the leader node being its root.
\end{assumption}

\begin{remark}
Note that Assumption \ref{AssumptionSpanningTree} is not restrictive. For
example, it holds when the subgraph describing the communication topology
between the followers is connected, and also at least one follower
senses the output of the leader. More generally, when the communication
topology between the followers consists of $p$ separate connected
components, Assumption \ref{AssumptionSpanningTree} will be satisfied if
each component of the graph includes a node which
directly senses the output of the leader.
\end{remark}

\subsection{The leader-following $H_\infty$ tracking problem}

The control problem in this paper is to design a distributed protocol
$u_{i}(k)$, $i=2,3,\cdots,N$, to enable the closed-loop multi-agent
system (\ref{FollowerDynamics}), equipped with this protocol, to
achieve a prescribed level of $H_{\infty}$ leader-following tracking
performance. The mathematical definition of the $H_{\infty}$ leader-following tracking
performance index will be given later. To guarantee the $H_{\infty}$ leader-following consensus
tracking performance, the following observer-based dynamic distributed
tracking protocol is proposed for each follower $i$, $i=2,3,\cdots,N$. The
protocol consists of two parts: 
 \begin{itemize}
\item [i)] The neighbor-based local controller:
\begin{equation}\label{LocalController}
u_{i}(k)=\check{A}x_{i}(k)
+\sum_{j=1}^{n-1}\hat{A}_{nj}z_{i}^{j}(k)-F_{n}\varepsilon_{i}(k),
\end{equation}
where $\check{A}=\hat{A}_{nn}-A$,  $\hat{A}_{ij}$, $i,j=1,2,\cdots,n$, are defined in (\ref{partition}) and $\varepsilon_{i}(k)$
is given in (\ref{RelativeInformation}).

\item [ii)] Distributed state estimator:
\begin{eqnarray}\label{DistributedEstimator}
z_{i}^{s}(k+1)=\hat{A}_{sn}x_{i}(k)+\sum_{j=1}^{n-1}\hat{A}_{sj}z_{i}^{j}(k)-
F_{s}\varepsilon_{i}(k), \\
s=1,2,\cdots,n-1. \nonumber
\end{eqnarray}
 \end{itemize}
The gain matrix $F=(F_{1}^{T},F_{2}^{T},\cdots,F_{n}^{T})^{T}\in
\mathbb{R}^{nm_{0}\times m_{y}}$ is the design parameter of the protocol
which will be defined later. 

Combining equations (\ref{FollowerDynamics}),
(\ref{LocalController}) and (\ref{DistributedEstimator}) yields the
closed-loop system describing dynamics of each follower governed by the
proposed protocol:
 \begin{equation}\label{ZetaEvolution}
 \zeta_{i}(k+1)=\hat{A}\zeta_{i}(k)-F\varepsilon_{i}(k)+\hat{B}_{\omega}{\omega}_{i}(k),\quad  k \in \mathbb{N},
 \end{equation}
where
\begin{equation}
\zeta_{i}(k)=\left(z_{i}^{1}(k)^{T},z_{i}^{2}(k)^{T},\cdots,z_{i}^{n-1}(k)^{T},x_{i}(k)^{T}\right)^{T}\in
\mathbb{R}^{nm_{0}}
\label{closed-loop-state}
\end{equation}
is the state of the closed-loop system,
$i=2,3,\cdots,N$, and
$\hat{B}_{\omega}=\big(O_{m_{0}\times m_{\omega}}^{T},\cdots,O_{m_{0}\times m_{\omega}}^{T},$ $B_{\omega}^{T}\big)^{T}\in \mathbb{R}^{nm_{0}\times m_{\omega}}$.
Then, it is easy to see that the difference between the state of the leader
and the state of the $i$-th closed-loop system,
$\rho_{i}(k)=\zeta_{i}(k)-\Theta(k)$, satisfies the following equation
 \begin{equation}\label{ErrorDynamics}
 \rho_{i}(k+1)=\hat{A}\rho_{i}(k)- F\varepsilon_{i}(k)+\hat{B}_{\omega}{\omega}_{i}(k),
 \end{equation}
$i=2,3\cdots,N$.

To characterize performance of the proposed tracking protocol,
define the performance variable
$e(k)=\left(e_{2}(k)^{T},e_{3}(k)^{T},\cdots,e_{N}(k)^{T}\right)^{T}$ where
$e_{i}(k)=C\rho_{i}(k)$ where  $C\in \mathbb{R}^{nm_{1}\times nm_{0}}$ is
a given performance output matrix, $i=2,3,\cdots,N$.

 Using the expression for $\varepsilon_i(k)$ given in
 (\ref{RelativeInformation}), we have
 \begin{equation} \label{ErrorDynamics-2}
\left\{\!\begin{aligned}
       &\rho(k+1){=}\left\{\!\left(I_{N-1}\otimes \hat{A}\right)-\left[\left(I_{N{-}1}{-}\breve{D}\right)\otimes (F\tilde{C})\right]\!\right\}\rho(k) \\
 &\quad \quad \quad \quad \;\;+
 (I_{N-1}\otimes \hat{B}_{\omega})\omega(k), \\
 &e(k+1)=\left(I_{N-1}\otimes C\right)\rho(k+1),
                          \end{aligned} \right.
                          \end{equation}
 where $\breve{D}$ denotes the $(N-1)\times (N-1)$ matrix defined as
 $\breve{D}=[\breve{d}_{ij}]_{(N-1)\times (N-1)}$ with
 $\breve{d}_{ii}={(h+\delta_{i})}/{\kappa}$,
 $\breve{d}_{ij}={a_{(i+1)(j+1)}}/{\kappa}$,
 $\omega(k)=({\omega}_{2}(k)^{T},{\omega}_{3}(k)^{T},\cdots,{\omega}_{N}(k)^{T})^{T}$ and
 $\delta_{i}=\kappa_{0}-\sum_{j=1}^{N}a_{(i+1)j}$, $\kappa$ and $\kappa_{0}$ are the constants defined in (\ref{RelativeInformation}), $\tilde{C}=(O_{m_{y}},\cdots,O_{m_{y}},\,E)\in
\mathbb{R}^{m_{y}\times nm_{0}}$ where $E$ is the matrix from equation
(\ref{OutputInformation}).
Denote by
$T_{\omega e}(\mathrm{z})$ the transfer function matrix of the system
(\ref{ErrorDynamics-2}) from disturbance input $\omega(k)$ to the
performance output $e(k)$ .

We are now in a position to formulate the leader-following $H_\infty$ 
tracking problem under consideration in this paper. 

 \begin{definition}\label{Definition1}
The multi-agent system consisting of the leader (\ref{LeaderDynamics}) and the
followers (\ref{FollowerDynamics}) and equipped with the protocol
(\ref{LocalController}) is said to solve the distributed $H_{\infty}$
leader-following
tracking problem with  performance index $\gamma>0$, if the following two conditions hold:
\begin{enumerate}
  \item [i)] The multi-agent system described by (\ref{LeaderDynamics}) and
    (\ref{FollowerDynamics}) with $\omega_{i}(k)\equiv 0$,
    $i=2,3,\cdots,N$,  achieves consensus in the sense of
    $\lim\limits_{k\rightarrow \infty}\|\zeta_{i}(k)-\Theta(k)\|=0$, where
    $\zeta_{i}$ is the state of the closed loop system defined in
    (\ref{closed-loop-state}),
    $i=2,3,\cdots,N$.

  \item [ii)]The $H_{\infty}$ norm of $T_{\omega e}(\mathrm{z})$ satisfies the following condition:
  $\|T_{\omega e}(\mathrm{z})\|_{\infty}<\gamma$.
\end{enumerate}
\end{definition}

\section{Main results}
In this section, the main theoretical results are presented.

 Since the leader has no neighbors,
 the matrix $D$ associated with communication topology
 $\mathcal{G}$, has the following structure
 \begin{eqnarray}\label{Row-StochasticD}
D&=&\left(\begin{array}{cc}
1&0\\
\breve{d}& \breve{D}
\end{array}\right), \\
\breve{d}&=&
(a_{21}/\kappa,a_{31}/\kappa,\cdots,a_{N1}/\kappa)^{T}\in
\mathbb{R}^{N-1}, \nonumber
\end{eqnarray}
and is a row-stochastic matrix; $\kappa$ is the constant defined in
(\ref{RelativeInformation}).
By Assumption \ref{AssumptionUndirected}, the
block $\breve{D}$ in (\ref{Row-StochasticD}) is symmetric. Let $\lambda_{i}$, $i=1,2,\cdots,N-1$, be the eigenvalues of $\breve{D}$.
It then follows from Lemma \ref{LemmaRowSto} and Assumption \ref{AssumptionSpanningTree} that
 $0<\lambda_{i}<1$, for all $i=1,2,\cdots,N-1$.

\begin{theorem}\label{MainTheorem1}
Suppose the communication
graph $\mathcal{G}$ satisfies Assumptions \ref{AssumptionUndirected} and  \ref{AssumptionSpanningTree}.
Then, for a given $\gamma>0$, the distributed $H_{\infty}$ leader-following
tracking problem stated in Definition \ref{Definition1} admits a solution
if and only if the following
$N-1$ systems are simultaneously internally stable and have the
$H_{\infty}$ norm less than $\gamma$:
\begin{equation} \label{MainTheorem-EQ1}
\left\{\begin{aligned}
       &\tilde{\rho}_{i}(k+1)=(\hat{A}-(1-\lambda_{i})F\tilde{C})\tilde{\rho}_{i}(k)+
 \hat{B}_{\omega}\tilde{\omega}_{i}(k), \\
 &\tilde{e}_{i}(k+1)=C\tilde{\rho}_{i}(k+1),
                          \end{aligned} \right.
                          \end{equation}
 where $i=1,2,\cdots,N-1$.
\end{theorem}

\begin{remark}\label{RemarkBelowTheorem1}
Theorem \ref{MainTheorem1} shows that the distributed $H_{\infty}$ tracking
problem for the networked agent system (\ref{ErrorDynamics-2}) can been
converted into a collection of $H_{\infty}$ control problems for a group of
uncoupled systems
(\ref{MainTheorem-EQ1}), each having the same dimension. Thus, the
complexity of the design reduces significantly. Note also that the effect
of topology on the distributed $H_{\infty}$ leader-following tracking
performance is characterized by the eigenvalues of the $\breve{D}$.
\end{remark}

Although Theorem \ref{MainTheorem1} gives necessary and sufficient
conditions for the distributed $H_{\infty}$ leader-following tracking
problem to admit a solution, it does not explain how the feedback gain
matrix $F$ should be selected in order to obtain such a solution. The
following theorem shows that this issue can be addressed using
tools from the $H_{\infty}$ control theory, based on the result of Theorem
\ref{MainTheorem1}.

 \begin{theorem}\label{MainTheorem2}
Suppose that Assumptions \ref{AssumptionUndirected} and
\ref{AssumptionSpanningTree} hold, and let
$\lambda_{0}=\max_{i=1,2,\cdots,N-1}|\lambda_{i}|$.
Given a constant $\gamma>0$, suppose
there exist real matrices $P=P^{T}>0$, $V$ and a
positive scalar $\varepsilon>0$ such  that
\begin{equation}\label{MainTheorem2FirstLMI}
\small{\left(\!\!\!\!\begin{array}{cccc}
-P\;\;&P\hat{A}{-}V\tilde{C}\;\;&\;P\hat{B}_{\omega}&\;V\\
\hat{A}^{T}\!\!P{-}\tilde{C}^{T}V^{T}\;\;&-P\!{+}\frac{1}{\gamma^2}C^{T}C+\varepsilon \lambda_{0}^{2}\tilde{C}^{T}\!\!\tilde{C}\;&O&\;O\\
\hat{B}_{\omega}^{T}P\;\;&O&-I\;&\;O\\
V^{T}\;\;&O&O&-\varepsilon I
\end{array}\!\!\right)\!\!<0,}
\end{equation}
and
\begin{equation}\label{MainTheorem2LMI2}
\left(\begin{array}{cc}
I & C\\
C^{T} & \gamma^{2}P
\end{array}
\right)>0.
\end{equation}
Then the protocol (\ref{LocalController}) augmented with the distributed
state estimator (\ref{DistributedEstimator}), with the feedback gain matrix
$F$, defined as $F=P^{-1}V$, solves the $H_{\infty}$
leader-following tracking problem for the multi-agent system described by
(\ref{LeaderDynamics}) and (\ref{FollowerDynamics}), with a disturbance
attenuation level $\gamma$.
\end{theorem}

\begin{remark}
Theorem~ \ref{MainTheorem2} provides sufficient conditions for
solvability of the distributed $H_{\infty}$ leader-following tracking
problem for the multi-agent system described by (\ref{LeaderDynamics}) and
(\ref{FollowerDynamics}). It is not hard to see that a necessary condition
for this tracking problem to have a solution is that the matrix
pair $(\tilde{C},\hat{A})$ must be detectable.
\end{remark}

\begin{remark}
Note that performance of the proposed $H_\infty$ tracking protocol is
determined by the matrix $C\in \mathbb{R}^{nm_{1}\times nm_{0}}$ in
(\ref{ErrorDynamics-2}) which defines the
performance variable $e(k)$. In the special case where the performance variable
of interest is the tracking error $\rho(k)$, the conditions of Theorem
\ref{MainTheorem2} are simplified by letting $C=I_{nm_{0}}$.
\end{remark}

\section{Example}
\begin{figure}[thpb]
      \centering
      \includegraphics[scale=1.0]{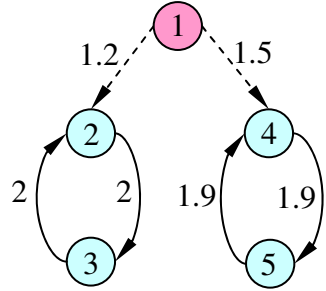}
      \caption{The communication topology $\mathcal{G}$.}
      \label{fig1}
   \end{figure}

Consider a leader whose dynamics are described by equation
(\ref{LeaderDynamics}), with
\begin{align*}\label{SimulationLeader}
\Theta(k)=\left(\!\!\begin{array}{c}
\theta_{1}(k)\\
\theta_{2}(k)\\
\theta_{3}(k)
\end{array}\!\!\right), \quad \hat{A}=\left(\!\!\begin{array}{ccc}
1&0&0\\
1&1&0\\
1&1&0.5
\end{array}\!\!\right),\quad \hat{C}=(0,0,1).
\end{align*}
Furthermore, let $m_{0}=1$, $E=1$, and $B_{\omega}=1.5$. Then the
corresponding matrices $\tilde{C}$ and $\hat{B}_{\omega}$ are
$\tilde{C}=(0,0,1)$, $\hat{B}_{\omega}=(0,0,1.5)^{T}$. Clearly, the pair $(\tilde{C},\,\hat{A})$ is detectable.  To illustrate
Theorem~\ref{MainTheorem2}, let us consider a network of agents of the form
(\ref{FollowerDynamics}) connected over the
communication graph $\mathcal{G}$ shown in Fig.~\ref{fig1}. The adjacency matrix of this graph is
\begin{equation*}\label{SimulationAdjacencyMatrix}
\mathcal{A}=\left(\begin{array}{cccccc}
0\;\;  &0\;\;\;&0\;&0\;&0\\
1.2\;\;&0\;\;\;&2\;&0\;&0\\
0\;\;  &2\;\;\;&0\;&0\;&0\\
1.5\;\;&0\;\;\;&0\;&0\;&{1.9}\\
0\;\;  &0\;\;\;&0\;&{1.9}\;&0
\end{array}\right).
\end{equation*}

It is easy to check that Assumption \ref{AssumptionSpanningTree}
holds. Thus, the neighbor-based protocol consisting of the local controller 
of the form (\ref{LocalController}) and the distributed 
 state estimator of the form (\ref{DistributedEstimator}) can be
 designed by solving the conditions in Theorem~\ref{MainTheorem2}.
 To design the protocol in this example, the parameter $h$ in
 (\ref{RelativeInformation}) is set
 to be equal $0.20$. Calculations show that the eigenvalues of $\breve{D}$
 defined in (\ref{Row-StochasticD}) are $\lambda_{1}=-0.2857$,
 $\lambda_{2}= -0.2844$, $\lambda_{3}=  0.8336$,
 $\lambda_{4}=  0.8597$. Then,
 $\lambda_{0}=\max_{i=1,\cdots,4}|\lambda_{i}|=0.8579$. Also,
 the output matrix $C=0.15I_{3}$ and performance level $\gamma=1$ were
 chosen in this example.

Solving the linear matrix inequality (\ref{MainTheorem2FirstLMI}) with
$\varepsilon=0.25$ gives that
$V=(
   -0.8698,\,
    0.2105,\,
    0.1083)$ and
$F=(    0.0003,\,
    0.0551,\,
    0.4660)^{T}$. To illustrate properties of this
protocol, we simulated the closed loop system without disturbances and also
with disturbance inputs of the form
 $\omega_{i}(k)=25\mathrm{sin}(i(k-1))\overline{\omega}(k)$, where
 $i=2,\cdots,5$, $k\in \mathbb{N}$, and
 \begin{equation}\label{DisturbanceInSimulation}
 \overline{\omega}(k)=\bigg\{\!\!\begin{array}{c}
1 \quad 0\leq k \leq 200,\\
0 \quad \mathrm{otherwise}.
\end{array}
 \end{equation}

\begin{figure}[h]
      \centering
      \includegraphics[scale=0.55]{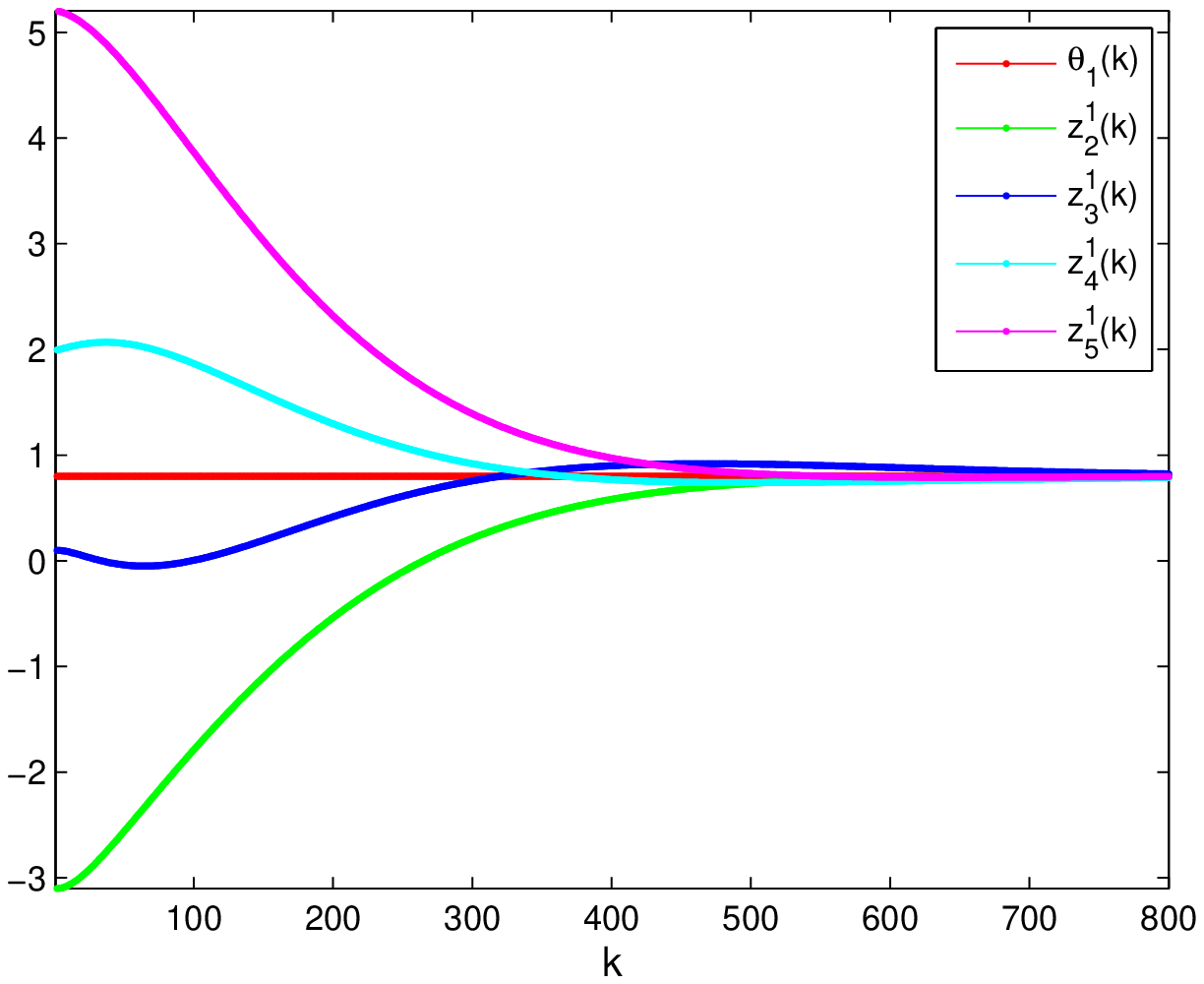}
      \caption{Trajectories of the leader's first state variable
        $\theta_{1}(k)$ and the corresponding estimate of this variable,
        $z_{i}^{1}(k)$, produced by the estimator embedded in the $i$th
        follower, $i=2,3,4,5$.}
      \label{figure2}
   \end{figure}

\begin{figure}[h]
      \centering
      \includegraphics[scale=0.55]{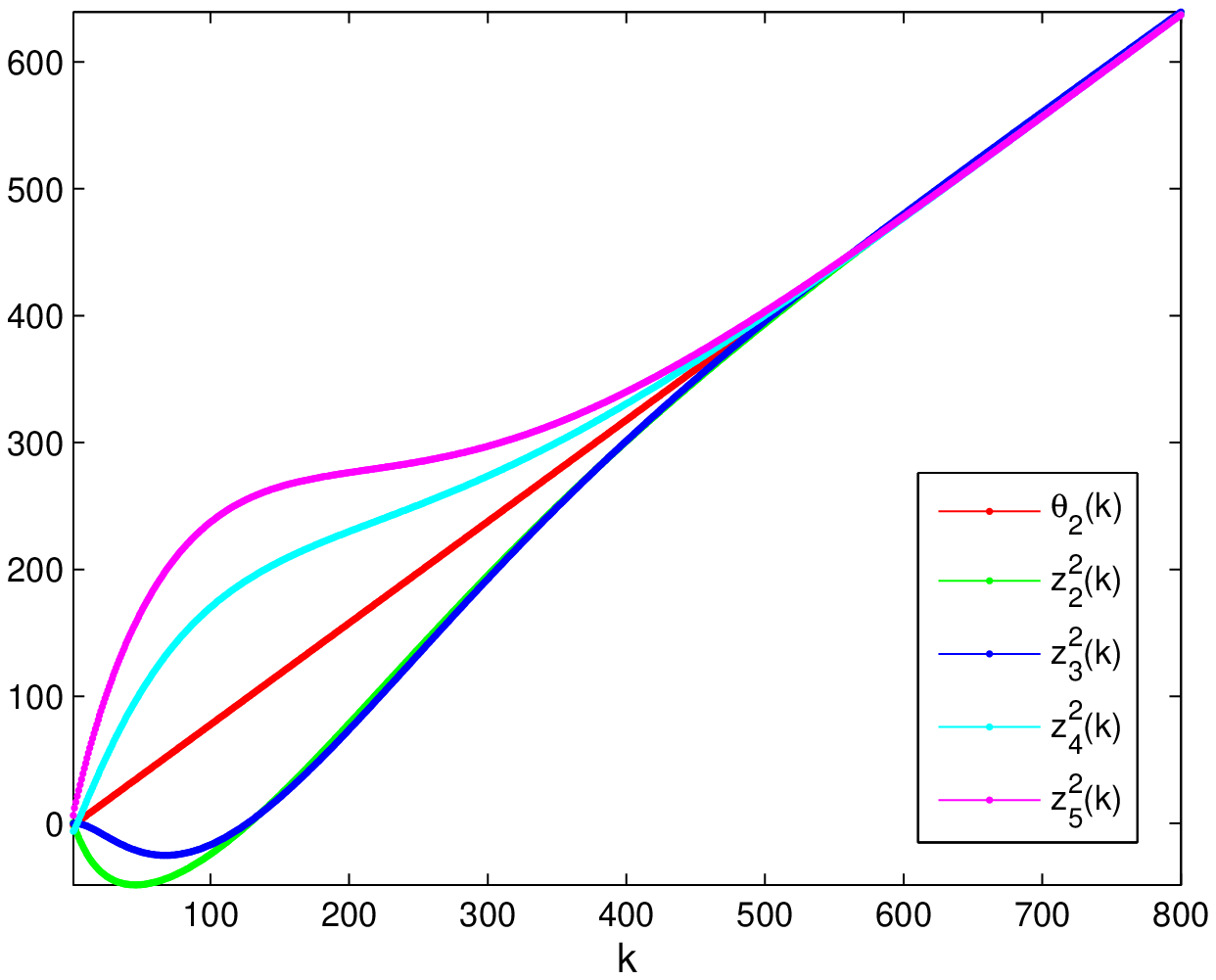}
      \caption{Trajectories of the leader's second state variable $\theta_{2}(k)$ and the corresponding estimate of this variable,
        $z_{i}^{2}(k)$, produced by the estimator embedded in the $i$th follower, $i=2,3,4,5$.}
      \label{figure3}
   \end{figure}

\begin{figure}[h]
      \centering
      \includegraphics[scale=0.55]{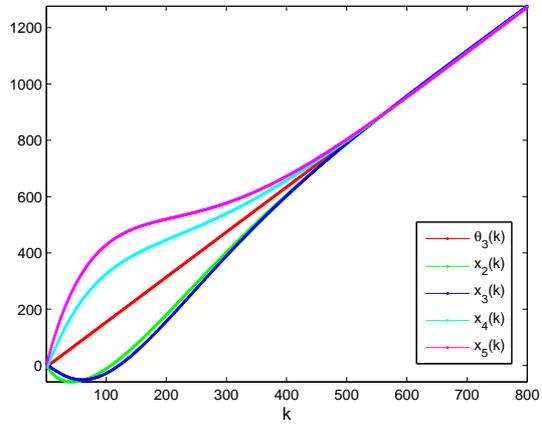}
      \caption{Trajectories of the leader's third state variable $\theta_{3}(k)$ and the followers' state variables $x_{i}(k)$, $i=2,3,4,5$.}
      \label{figure4}
   \end{figure}

\begin{figure}[h]
      \centering
      \includegraphics[scale=0.55]{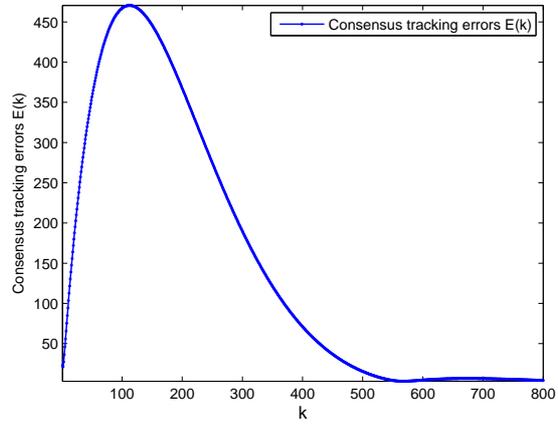}
      \caption{The squared norm of the consensus tracking error $E(k)$, in
        the absence of disturbances.}
      \label{figure5}
   \end{figure}

\begin{figure}[h]
      \centering
      \includegraphics[scale=0.55]{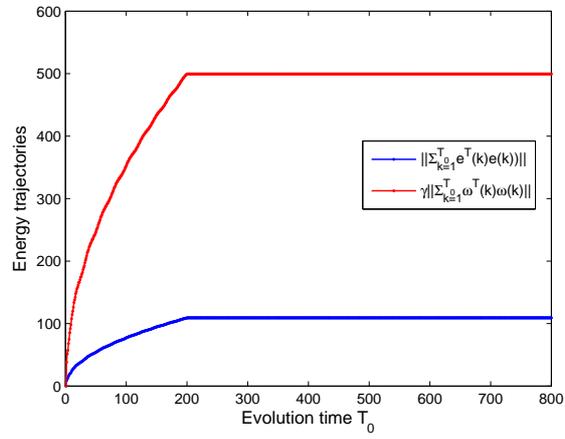}
      \caption{Energy trajectories of the
          performance output $e(k)$ (the bottom curve) and the
        disturbance $\omega(k)$.}
      \label{figure9}
   \end{figure}

To illustrate asymptotic convergence of the tracking agents in the absence
of disturbances, the corresponding state trajectories of the closed-loop
multi-agent system
(\ref{FollowerDynamics}) with a high-dimensional leader
(\ref{LeaderDynamics}), are shown in Figs. \ref{figure2}--\ref{figure4}.
Let $E(k)=\sum_{j=2}^{5}\|\zeta_{j}(k)-\theta(k)\|^2$ be the
square of the norm of the consensus tracking error vector for the multi-agent system, where
$\zeta_{j}(k)=(z_{j}^{1}(k),z_{j}^{2}(k),x_{j}(k))^{T}$,
$j=2,\cdots,5$. Fig.~\ref{figure5} indicates that the proposed
distributed dynamic tracking protocol indeed ensures
consensus tracking in the absence of disturbances, i.e., when
$\omega_{i}(k)\equiv 0$,  $i=2,\cdots,5$. Next, the $H_{\infty}$ consensus
tracking under disturbances is considered. Under zero initial
conditions, the `energy trajectories' $\|\sum_{k=1}^{T_{0}}e^{T}(k)e(k)\|$ and
$\gamma\|\sum_{k=1}^{T_{0}}\omega^{T}(k)\omega(k)\|$ were computed  as
functions of the evolution time $T_{0}$  and were
plotted in Fig.~\ref{figure9}. It can be seen from Fig.~\ref{figure9} that the
proposed distributed dynamic tracking protocol indeed ensures the set level of
disturbance attenuation.

\section{Conclusions}

The distributed $H_{\infty}$ leader-following tracking problem for a class of
discrete-time multi-agent systems with a high-dimensional active leader has been investigated in this paper. In the
presented framework, the outputs of the leader are only sensed by some
informed followers. A new kind of dynamic tracking protocol consisting of a
local controller and a distributed state estimator has been constructed and
employed to solve such a coordination problem. Using tools from the $H_{\infty}$
control theory, it has been proved that distributed $H_{\infty}$
leader-following tracking can be ensured if the underlying topology graph
contains a directed spanning tree with the leader being its root while the
communication topology among the followers is undirected. Future work will
focus on solving the distributed 
tracking problem for multi-agent systems with a high-dimensional leader and
time-varying topologies as well as leader-following tracking for multi-agent
systems with nonlinear dynamics.

\addtolength{\textheight}{-12cm}   




\section*{ACKNOWLEDGMENT}
G. Wen would like to thank Prof. Yiguang Hong for the inspiring discussions
and helpful suggestions.

\newpage

\end{document}